\documentclass[a4paper,11pt]{article}
\usepackage{pos}

\title{The ubiquitous mechanism accelerating cosmic rays at all the energies}
\author*[a]{Antonio Codino}
\affiliation[a]{University of Perugia and INFN,\\
  Via A. Pascoli, Perugia, Italy}

\emailAdd{antonio.codino@pg.infn.it}
\abstract{The mechanism accelerating cosmic rays in the M ilky W ay Galaxy and
galaxy clusters is identified and described. The acceleration of cosmic rays is
a purely electrostatic process which operates up to the maximum energies of
$10^{23}$ $eV$ in galaxy clusters. Galactic cosmic rays are accelerated in a pervasive electrostatic field active in the whole Galaxy except in restricted regions 
shielded by interstellar and stellar plasmas as, for instance, the region occupied
 by the solar system.\\
It is proved that the energy spectrum of the
 cosmic radiation in the $Milky$ $Way$ $Galaxy$, in the region where the
 solar system resides, has a constant spectral index comprised between
 2.64-2.68 and the maximum energies of Galactic protons are $3.0 \times
10^{19}$ $eV$.  The agreement of these results with the experimental
data is discussed in detail and underlined.\\

 The various physical processes that maintain the stability of the
 electrostatic  structure in the $Milky$ $Way$ $Galaxy$ are the same
 that generate the Galactic magnetic field.
 Accordingly, the
 intensity, orientation and direction of the Galactic magnetic field
 are evaluated. The results of the calculation are compared
 with the observational data, optical and mostly radio astronomy data. The
accord of the intensity, orientation and direction of the observed
magnetic field with calculation is excellent.}

\FullConference{37$^{\rm{th}}$ International Cosmic Ray Conference (ICRC 2021)\\
		July 12th -- 23rd, 2021\\
		Online -- Berlin, Germany}

\begin{document}
\maketitle

\section{Introduction}
\quad Experimental and theoretical studies have attempted for many
decades to identify the mechanisms and cosmic sites where the atomic
nuclei of the quiescent galactic matter are converted into cosmic
rays, namely,  ions of ultrarelativistic energies. Particle
acceleration mechanisms have been conceived in 1954 by Enrico Fermi
\cite{fermi1, fermi2} and have survived, with a number of variants,  until the
present days.

\quad Such mechanisms are not adequate to account for the
experimental data and particularly, the differential energy spectrum
of the cosmic radiation, $J$, which extends up to the energy of $3
\times 10^{20}$ $eV$ \cite{agasa, fly, ta, verzi}. This spectrum as it is well known, in
restricted energy bands, can be expressed by a piecewise power law,
 $J$ = $a$/$E^{\gamma}$ \quad where $E$ is the cosmic ray energy,
$\gamma$ is the spectral index and $a$ is a constant. For example,
in the interval $10^{10}$-$2 \times 10^{15}$ $eV$ from a variety of
experimental data it turns out, $a$ = 9574 particles/$m^2$ $s$ $sr$
$GeV^{1.67}$, an index $\gamma$ of 2.67 \cite{tracer} and a particle rate of
\quad one particle/$m^2$ $s$ $sr$ $GeV$ \quad at the energy of
$30.9$ $GeV$. At this arbitrary energy the intensity is not reduced
by the solar modulation.

\quad  A new and universal mechanism is
proposed in this paper which explains the acceleration of cosmic
rays from the energy of the quiescent matter up to $10^{23}$ $eV$
and beyond. Energies of the quiescent matter are those of the
ionization of atomic nuclei, crudely in the range $5$ $eV$-$10$
$keV$ along with those of cold interstellar matter down to energies
of $10^{-3}$ $eV$.

\section{The electrostatic field residing in the Milky Way Galaxy}
he observed energies of extrasolar cosmic rays extend from
about $30$ $MeV$ to $2 \times 10^{20}$ $eV$ \cite{agasa, fly, ta, verzi}. The observed
spectral index $\gamma$ is measured to be in the range 2.6-2.7 for
all cosmic nuclei with energies below $10^{15}$ $eV$. Once the
energy spectrum is known,  the total kinetic energy transported by
the cosmic rays can be determined. The kinetic energy  per unit
volume turns out to be about 1.1 $eV$/$cm^3$. Notice that in the
evaluation of the energy density of cosmic rays the maximum energy
of the spectrum is not critical since the major fraction of the
energy is concentrated between  0.5 and 10 $GeV$. Accordingly, the
energy of cosmic rays per unit volume is an observed, empirical
quantity, at least a secure order of magnitude.

\quad Cosmic rays are charged particles and, therefore, the
simplest, conceivable  means they acquire the enormous amount of
kinetic energy they store might originate from an electric field. In
this case the problem to resolve is to recognize where this electric
field is located,  how it is sometimes hidden in the cosmic ambient
and how the acceleration of nuclei, electrons, positron and
antiprotons would take place.

\section{The total negative charge stored in spiral galaxies}
If the kinetic energy of cosmic rays
would originate from an electrostatic source,  the order of
magnitude of the amount of electric charge which would generate the
electrostatic potential  $V_e$ and the related electrostatic energy
$U_e$ can be determined by the $Virial$ $Theorem$. As cosmic rays
are a population of charged particles, when they are immersed in an
electrostatic field they respond via the Coulomb law which depends
on the square of the particle distance between particle pairs. This
theorem dictates that after sufficiently long time, the average
kinetic energy  $\overline T_{cr}$ of a system of particles endowed
with electric charge  equals the half of the electrostatic potential
energy, $\overline U_e$,  namely :

$$  \overline U_e  =  - 2  \overline T_{cr}  \eqno (1)$$

For sake of simplicity let us imagine a sphere of radius
$r_g$= $15$ $kpc$ and a volume  $V$ = 4/3 $\pi$ $r^3_g$ = $415
\times$ $10^{66}$ $cm^3$ representative of the volume of a typical
galaxy like the $Milky$ $Way$ $Galaxy$ albeit this celestial body is
not spherical. Let $Q_{cr}$ be the electric charge of cosmic rays
which is globally positive, and it is carried by cosmic nuclei. The
electric charge conservation in a finite volume requires an equal
amount of negative charge in the Galactic volume denoted here by
$Q_w$, and consequently, the following relation holds: \quad $Q_{cr}$ 
+ $Q_w$ = $0$ \quad regardless of the way cosmic rays are
accelerated. The negative charge  $Q_w$ ( $w$  is for widow
electron) is stored by electrons of 
the quiescent Galactic matter being these particles the only stable 
and available.

\quad Preliminarily,  let us admit that the negative electric charge
$Q_w$ is uniformly distributed in the spherical volume of radius
$r_g$ while the positive charge  $Q_{cr}$ has a spherical symmetry
with a decreasing radial density of the form $k$/$r$ appropriate for
cosmic rays emanating from a finite source where $k$ is an
appropriate constant. The dependence $1$/$r$ is a characteristic
feature of the diffusive motion and it extends beyond $r_g$ up to a
maximum distance $R$ (for instance $R$ = 20 $r_g$). Fig. 1
highlights  what has been imagined up here showing, qualitatively,
the positions of the negative and positive electric charges of
cosmic rays while neutral matter is omitted for simplicity.

\quad In the specified conditions, the total electric charge $q_g$
within the sphere of radius  $r_g$ is : $q_g$ = $\lbrack$ $Q_w$ -
$Q_{cr}$($r_g$) $\rbrack$ = $Q_w$ ($r_g$/$R)^2$. The electrostatic
energy  $E_e$ contained in the same sphere amounts to: $E_e$ =
$Q^2_{w}$ /($4$ $\pi$ $\epsilon_0$ $R$ )(-$l/4$ +$l^3$/$6$ +
$l^5$/$10$) \quad being $l$ $\equiv$ ($r_g$/$R$). Since $l$  is a
small number,  the terms ($l^3$/$6$ + $l^5$/$10$) can be neglected
and it results : \quad $E_e$ = $Q^2_{w}$ $r_g$ /($16$ $\pi$
$\epsilon_0$ $R^2$ ). \quad A first good approximation is to apply
equation (1) to  the volume $V$ and the result is: \quad $\overline
U_e$$V$ = 2 $\overline T_{cr}$ $V$ = $E_e$ \quad and hence the
charge  $q_g$ is:
$$ q_g =  \pi r_g R \lbrack {(128/3) \epsilon_0
 \overline T_{cr} } \rbrack^{1/2}  \eqno (2)$$
Substituting  the values of $r_g$, $R$ and $\overline T_{cr}$ in
equation (2) the resulting charge is  $q_g$ = $1.09$ $10^{32}$
$Coulomb$ (hereafter $C$) \samepage{\footnote{\it The analytical
calculation of the electric field in a sphere of volume $V$ is
straightforward and partially justifies the adoption of this
geometrical form. A disk-shaped volume with half thickness $h$ =
$125$ $pc$ and radius $r_g$ = $15$ $kpc$, a more appropriate
surrogate of the disk volume of the $Milky$ $Way$ $Galaxy$ yields an
electric charge  $q_g$ comparable to that obtained with the equation
(2). Detailed calculation of the electrostatic field and related
potential in a flat disk, surrounded by a positively charged larger
halo, are given in $Chap.$ $9$ of ref. \cite{codino20book}.\rm}}.
\begin{figure}
\begin{center}
  \includegraphics[width=10cm]{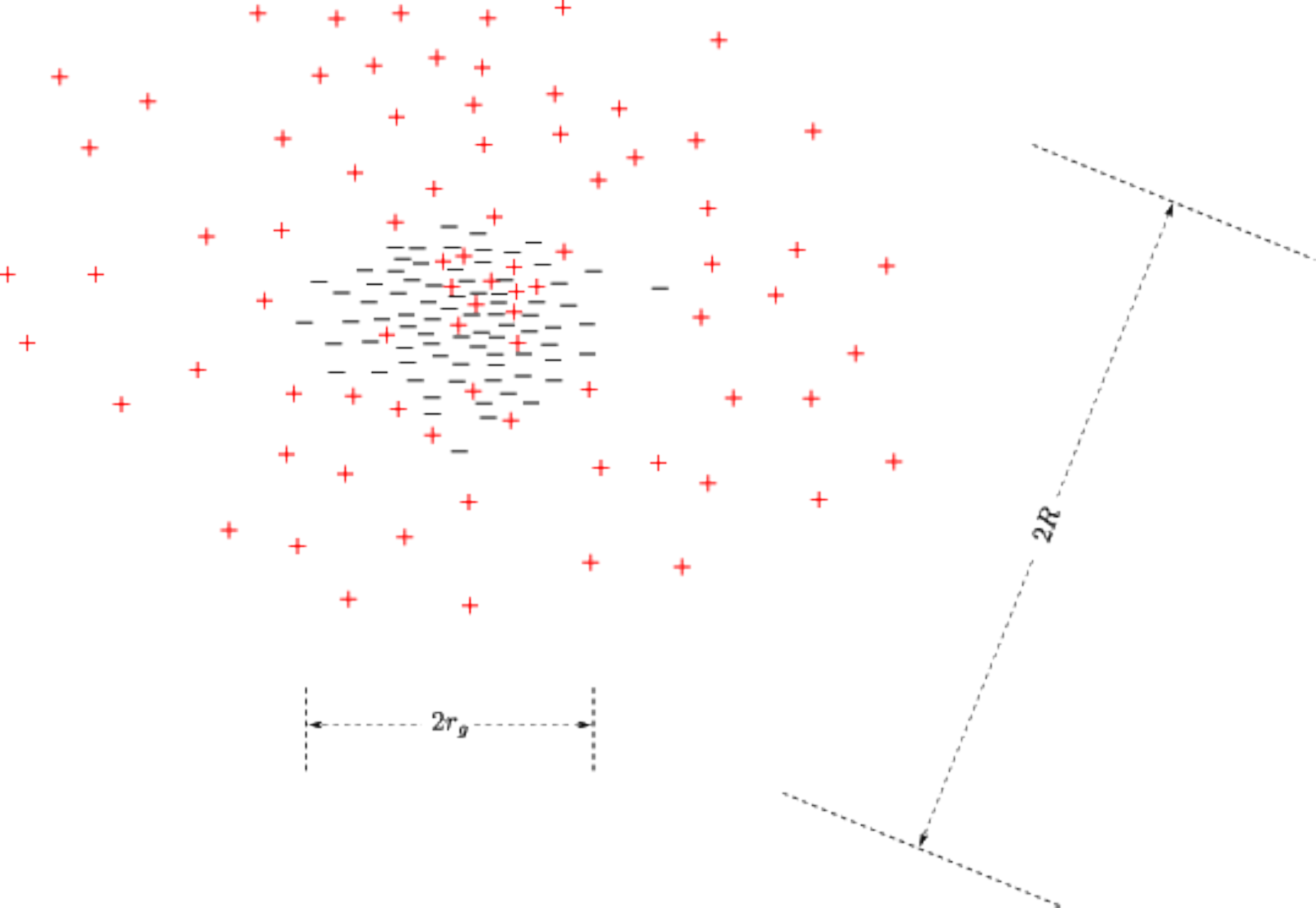}
  \caption{Qualitative illustration of the positions of the electric charges
transported by cosmic rays in a spherical globe of maximum size
$2R$. \quad The negative charges of the electrons of the quiescent
galactic matter are represented by negative signs and they are
confined in a region of size $2$$r_g$ while the positive charges of
protons, nuclei and positrons represented by plus signs extend up to
$2R$. \quad High energy electrons, unlike protons and nuclei,
radiate synchrotron (curvature) light and cannot diffuse over large
distances from the globe of characteristic size $2$$r_g$. Cosmic
nuclei propagate via a diffusive-alternate motion (see Appendix
$B.3$ and fig. 81 of ref. \cite{codino20book}) forming a halo of positive charge whose density 
decreases with the radial distance $r$ in the range $0$ $\le$  $r$ 
$\le$ $R$. The different distributions of the negative  (quiescent 
electrons) and positive (cosmic nuclei) charges generate an 
electrostatic field denoted $\vec E_g$. The cartoon is not at 
scale and ignores many details discussed  elsewhere \cite{codino20book}.}
\end{center}
\end{figure}

\section{On the intensity of the radial magnetic field  of the Milky Way Galaxy}

\begin{figure}
\begin{center}
  \includegraphics[width=10cm]{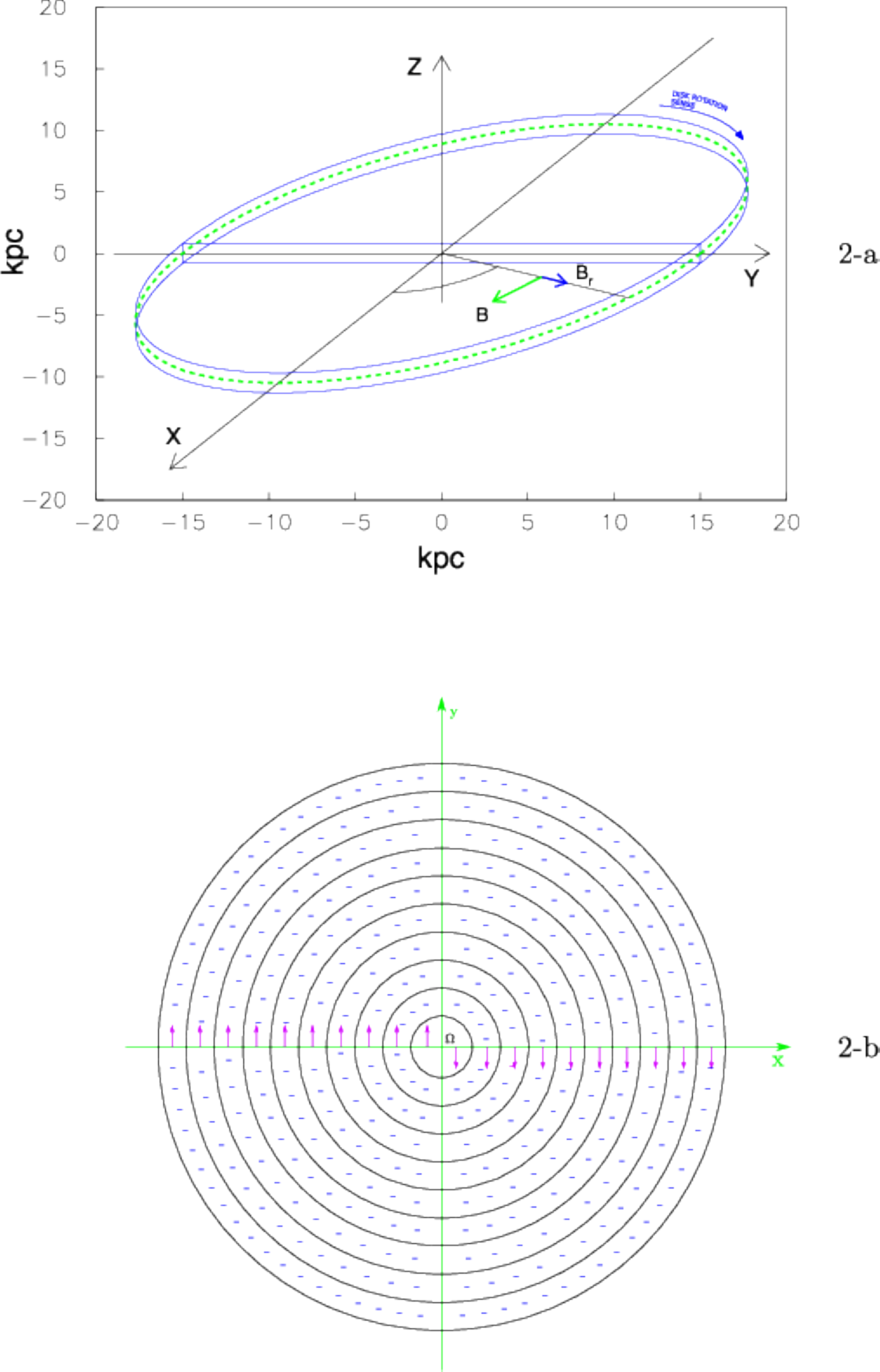}
  \caption{Illustration of a magnetic field  generated by a rigid disk rotating
about the  $z$ axis at the velocity of 245 $Km$/$s$. The height $h$
of the disk is $200$ $pc$ and the  radius  $r_g$ is $15 000$ $pc$.
The negative charge $q_g$ of - $1.09 \times 10^{32}$ $C$ \quad is
uniformly distributed in the disk volume. \quad $2$-$a$ The
reference system in the disk adopted in this work, \quad the
orientation of the axes $x$ $y$ and $z$, \quad the sense of rotation
seen from the north Galactic pole (blue arrow), \quad and the
Galactic midplane (annular, green, dashed curve) are shown. It is
also indicated the orientation of the $\phi$ angle which defines the
azimuthal component $\vec B_{\phi}$ of the Galactic magnetic field.
\quad $2$-$b$ \quad The 10 concentric  coronas resulting from the
partition of the disk volume used in the calculation of the radial
magnetic field. The violet arrows indicate the velocity orientation
of the 10 coronas seen from above and the minus signs denote the
negative charge of the rotating quiescent electrons.}
\end{center}
\end{figure}
\quad Gas, dust and stars of the $Milky$ $Way$ $Galaxy$ regularly
rotate about the Galactic center with typical velocities in the
range $230$-$250$ $Km$/$s$. These velocities, as it is well known,
are measured with great precision. Measurements of the circular
velocities of $230$-$250$ $Km$/$s$ of clouds and stars  span the distances  $300$-$400$ $pc$ 
from the Galactic center up to $15000$ $pc$ and beyond.

\quad An electric charge in a circular motion produces a
characteristic magnetic field. Let us assume that spiral galaxies
retain a negative electric charge of the same order of magnitude of
$q_g$ and simplify the form of the $Milky$ $Way$ $Galaxy$ by a
rotating rigid disk. Subdivide such a disk in $n$ circular coronas,
as shown in fig. 2-b. Any corona is endowed with an equal amount of
electric charge $q_n$ = $q_g$/$n$. In these conditions the order of
magnitude of the magnetic field originated by the circular motion
can be straightforwardly estimated.

\quad The magnetic field strength at the center of the circular coil
of radius $r_n$ traversed by the current $i_n$ is, \quad $\mu_0$
$i_n$/$2$ $r_n$ \quad the direction of the magnetic field is normal
to the disk plane. The total field strength $\vec B$ results from
the sum of all currents $i_n$ in the  $n$ coils, \quad $\vec B$ =
$\sum_n$ $\vec B_n$ = ($\sum_n$ $\mu_0$ $i_n$/$2$ $r_n$ ) $\vec u_z$
\quad being $r_n$ the radius of the $n$-$th$ circular coil, $\vec
u_z$ the unit vector of the $z$ axis and $i_n$ = $q_n$/$T_n$ with
$T_n$ the revolution period of the electric charge $q_n$ around the
Galactic center. For instance, with $n$ = 10, charges $q_n$
proportional to the coil volumes, equal velocities of 245 $Km$/$s$
for all the coronas which are placed at the radial distances ($r_n$
- $r_{n-1}$)/2, it turns out a magnetic field strength of 1.04
$\mu$$G$.

\quad The order of magnitude of such a field strength  equals that
of the regular magnetic field measured in the $Milky$ $Way$
$Galaxy$, in $Andromeda$ and in many other spiral galaxies. A
precise evaluation of the intensity, orientation, and direction of
the Galactic magnetic field is reported in the magnetic chapters of
ref. \cite{codino20book}.

\quad In summary, taking advantage of the $Virial$ $Theorem$
expressed by equation (1), the density of kinetic energy of cosmic
rays per unit volume of 1.1 $eV$/$cm^3$ enables to determine the
amount of electric charge that generates the corresponding
electrostatic potential $V_e$($r$). The reasoning up to this point,
rests on a simple and natural assumption.  The main mode any electric
charges like cosmic rays undergo acceleration is by an electric
field.

\quad It has been realized that the magnetic field strengths in
spiral galaxies fall in the correct order of magnitude under the
assumption that the electric charge  $Q_w$ rotates around the
Galactic center, as calculated above. Thus, the assumption of an
electric charge $Q_w$ in the $Galaxy$ already introduced with
independent arguments ($Seg. $ $2.1$), deserves an adequate
exploration and justifies the quest for the mechanism accelerating
cosmic rays along the electrostatic avenue.

\section{Maximum energies of Galactic cosmic rays}
\quad Proceeding further, the same amount of charge $q$($r_g$) has
its distinctive electrostatic potential $V_e$($r$) which at the
surface of sphere $r$ = $r_g$ is $V_e$ ($r_g$) = $q_{g}$/4 $ \pi
\epsilon_0 r_g$ \quad and with $r_g$ = $15$ $kpc$ and $q_g$ = $1.097
\times 10^{32}$ $C$ gives $21.3 \times 10^{20}$ $V$. Let us denote
$V_e^{ma}$ the maximum value of the potential on the spherical
surface at  $r_g$. A cosmic proton of charge $q$ moving from the
peripheral zone, $r$ = $r_g$ \quad to the center of the sphere $r$ =
$0$ $pc$,  would gain the energy $q$ $V_e$($r_g$) which is
$21.3\times 10^{20}$ $eV$.

\quad The $Milky$ $Way$ $Galaxy$ is not spherical but disk shaped
and,  a more realistic halo of positive electric charge, for example
$R$ = $r_g$ = $15$ $kpc$ along with a  charge  $q_g$= $2.5 \times
10^{31}$ $C$ (evaluated with more precise geometrical Galactic parameters in $Chap.$ $9$ of ref. \cite{codino20book}),  slightly less than that
reckoned in $Seg.$ $2.1$, would give an electrostatic potential on
the disk rim at $r_g$ = $15$ $kpc$ of $-3.64 \times 10^{20}$ $V$. At
the radial distance of the $Sun$ at $r_s$ = $8.5$ $kpc$ the
potential is, $-5.40 \times 10^{20}$ $V$ ($Seg.$ $8.1$ of ref. \cite{codino20book}). In these
 conditions the maximum kinetic energies of protons are $1.7 \times 
 10^{20}$ $eV$.

\quad  It results that protons wander in the $Milky$ $Galaxy$ with
maximum energies in the range ($3$-$8$)$\times 10^{19}$ $eV$. This
energy interval is close to the maximum energies of cosmic rays
measured at $Earth$ of about $3\times 10^{20}$ $eV$ \cite{agasa, fly, ta, verzi}. These
energies coincide with the energy band ($2$-$6$)$\times 10^{19}$
$eV$ where a rapid flux drop (cutoff) relative to the extrapolated power-law
smooth spectrum has been observed \cite{abbasi}. The cutoff energy region has
been interpreted as unmistakable, distinctive signature of maximum
energies of Galactic protons (see $Chap.$ $13$ of ref. \cite{codino20book}). Thus, again
 experimental data \cite{agasa, fly, ta, verzi} consistently revisited for the correct energy scale \cite{codino13,codino20} are compatible with the crude but essential
 evaluations just mentioned.

\par \quad Besides the concordance in the order of magnitude
of the galactic magnetic field with that  derived in $Seg.$ $2.2$
based on the circular motion of the charge $q_g$ stored in the disc,
an additional, independent agreement of order of magnitude, emerges
from the comparison of the maximum energies of the cosmic rays and
the maximum electrostatic potential energy per particle \quad
$Z$$q$$V^{ma}_e$.

\quad The sentiment is that the above concordance in two disparate
areas is  not a mere, casual coincidence since  these two physical
quantities, \quad maximum energy of cosmic protons and the intensity
of the Galactic magnetic field \quad have very distinctive,
unmistakable values, e.g. $2.7 \times 10^{19}$ $eV$ and $10^{-10}$
$T$. These two incredibly disparate numerical figures communicate by
the electrostatic field resident in the $Galaxy$ if the intuition
conceived at the beginning of this work is correct. Let us remind
that the differential intensity of the cosmic radiation $J$($E$)
span 31 orders of magnitude over 10 energy decades  (see fig. 46 in
ref. \cite{codino20book}).

\section{Electrostatic field intensity in the Milky Way Galaxy}
As it is well known the electrostatic energy of a system of
electric charges can be regarded not only as the total work
necessary to displace the electric charges from very large distances
to the positions they actually occupy in the system but, also, as
energy density of the electrostatic field which permeates the space
around the charges. Accordingly, the average intensity of the
electrostatic field $\overline E_g$ in the spherical volume $V$ can
be crudely estimated by the equation :
$$  E_e  = {1 \over 2 V} \int  \epsilon_0 E^2_{g} dV  = { \epsilon_0  \over 2 V}
( \overline E_g) ^2  V  =  { \epsilon_0  \over 2} ( \overline E_g)
^2 \quad \eqno (4)$$ % \quad being $dV$ the infinitesimal volume.

%The intensity of the electrostatic field $\overline E_g$ in the
%spherical volume $V$ can be crudely estimated by the equation :
%$$  E_e  = {1 \over 2 V} \int  \epsilon_0 E^2_{g} dV  = { \epsilon_0  \over 2 V}
%( \overline E_g) ^2  V  =  { \epsilon_0  \over 2} ( \overline E_g)
%^2 \quad \eqno (4)$$
\quad being $dV$ the infinitesimal volume
element, $\vec E_g$ the electrostatic field at any point in space
and $\overline E_g$  its average value. The electrostatic energy
$E_e$ also follows from equation (1),  \quad $E_e$ = $2$ $\overline
T_{cr}$$V$ \quad which substituted in the first member of equation
(4) gives an estimate of the average value of the electrostatic
field. The result is :
$$  \overline E_{g}  = (4    \overline T_{cr} / \epsilon_0 )^{1/2}      \eqno (5)$$
giving $\overline E_{g}$  = $0.28$ $V$/$m$.

\quad Detailed calculation of the shape and intensity of the
Galactic magnetic field $\vec B_g$ requires the knowledge of the
dynamics of cosmic rays and quiescent charged particles in the
$Milky$ $Way$ $Galaxy$ and overwhelms the size of this conference paper. Similarly, the spectral index $\gamma$ of 2.65 of the
cosmic-ray spectrum is calculated elsewhere \cite{codino20book}.
Both the index $\gamma$ and $\vec{B_g}$ are found in agreement with the experimental data as evidenced in \cite{codino20book}.

\end{document}